\documentclass{osa-article}

\journal{oe}

\articletype{Research Article}

\usepackage{dcolumn}
\usepackage{bm}
\usepackage{graphicx}
\usepackage{xcolor}
\usepackage{url}
\usepackage[normalem]{ulem}
\usepackage{times}
\usepackage{comment}
\usepackage{braket}

\begin{document}
\title{Self-healing high-dimensional quantum key distribution using hybrid spin-orbit Bessel states}
  \author{Isaac Nape,\authormark{1} Eileen Otte,\authormark{2} Adam Vall\'es,\authormark{1,*} Carmelo~Rosales-Guzm\'an,\authormark{1}  Filippo Cardano,\authormark{3} Cornelia~Denz,\authormark{2} and Andrew Forbes\authormark{1}}
\address{\authormark{1}School of Physics, University of the Witwatersrand, Private Bag 3, Wits 2050, South Africa\\
\authormark{2}Institute of Applied Physics, University of Muenster, Corrensstr. 2/4, D-48149 Muenster, Germany\\
\authormark{3}Dipartimento di Fisica "Ettore Pancini", Universit\'a di Napoli Federico II, Complesso Universitario di Monte Sant'Angelo, Via Cinthia, 80126 Napoli, Italy}

\email{\authormark{*}adam.vallesmari@wits.ac.za}




\begin{abstract}

\noindent Using spatial modes for quantum key distribution (QKD) has become highly topical due to their infinite dimensionality, promising high information capacity per photon. However, spatial distortions reduce the feasible secret key rates and compromise the security of a quantum channel. In an extreme form such a distortion might be a physical obstacle, impeding line-of-sight for free-space channels. Here, by controlling the radial degree of freedom of a photon's spatial mode, we are able to demonstrate hybrid high-dimensional QKD through obstacles with self-reconstructing single photons. We construct high-dimensional mutually unbiased bases using spin-orbit hybrid states that are radially modulated with a non-diffracting Bessel-Gaussian (BG) profile, and show secure transmission through partially obstructed quantum links. Using a prepare-measure protocol we report higher quantum state self-reconstruction and information retention for the non-diffracting BG modes as compared to Laguerre-Gaussian modes, obtaining a quantum bit error rate (QBER) that is up to 3$\times$ lower. This work highlights the importance of controlling the radial mode of single photons in quantum information processing and communication as well as the advantages of QKD with hybrid states.
\end{abstract}


\section{Introduction}
Quantum key distribution (QKD) enables two parties to securely exchange information detecting the presence of eavesdropping \cite{bennett1984quantum}. Unlike conventional cryptography, with unproven computational assumptions, the security of QKD relies on the fundamental laws of quantum mechanics \cite{shor2000simple}, prohibiting the cloning of quantum information encoded in single photons \cite{wootters1982single}. Although current state of the art implementations have successfully transfered quantum states in free-space \cite{schmitt2007experimental}, optical fibers \cite{gobby2004quantum}, and between satellites \cite{liao2017satellite}, efficient high capacity key generation and robust security are still highly sought-after.

Spatial modes of light hold significant promise in addressing these issues. The channel capacity can be exponentially increased by encoding information in the spatial degree of freedom (DoF) of photons and has been demonstrated with classical light in free-space and fibres \cite{bozinovic2013terabit}.  Implementing QKD with high-dimensional (HD) states ($d>2$) has also been demonstrated \cite{groblacher2006experimental, mafu2013higher}, by exploiting the ability of each photon to carry up to $\log_2(d)$ bits per photon while simultaneously increasing the threshold of the quantum bit error rate (QBER). This makes HD QKD protocols more robust \cite{bechmann2000quantum,cerf2002security, ali2007large}, even when considering extreme perturbing conditions, i.e., underwater submarine communication links \cite{bouchard2018underwater}.  While most studies to date have used spatial modes of light carrying orbital angular momentum (OAM) \cite{molina2007twisted}, reaching up to $d=7$ \cite{mirhosseini2015high}, higher dimensions are achievable with coupled spatial and polarization structures, e.g. vector modes.  These states have received recent attention in classical communication  \cite{Milione2015e,Li2016, Milione2015f,liu2018direct}, in the quantum realm as a means of implementing QKD without a reference frame \cite{Souza2008, vallone2014free} and for real-time quantum error correction \cite{ndagano2017hybrid}, but only recently have both DoFs been used to increase dimensionality in QKD \cite{Ndagano2018,ndagano2017det,sit2017high}.

To date, there has been only limited work on the impact of perturbations on HD entanglement and QKD with spatial modes \cite{huber2013weak, krenn2014communication, sit2017high, zhang2016,McLaren2014}.  In turbulence, for example, the key rates are known to decrease \cite{erven2012studying}, with the latter to be compensated for large OAM states in the superposition.  There has been no study on HD QKD through physical obstacles.

Here, we take advantage of the self-healing properties in non-diffracting vector beams to show that the bit rate of a QKD channel, affected by partial obstructions, can be ameliorated by encoding information onto diffraction-free single photons. To this end, we generate a non-diffracting (self-reconstructing) set of mutually unbiased bases (MUB), formed by hybrid scalar and vector modes with a Bessel-Gaussian (BG) transverse profile. We herald a single photon with a BG radial profile by means of spontaneous parametric down-conversion (SPDC), generating paired photons and coupling OAM and polarization using a $q$-plate \cite{Marrucci2006}. We characterize the quantum link by measuring the scattering probabilities, mutual information and secret key rates in a prepare-measure protocol for BG and Laguerre-Gaussian (LG) photons, comparing the two for various obstacle sizes. We find that the BG modes outperform LG modes for larger obstructions by more than $3\times$, highlighting the importance of radial mode control of single photons for quantum information processing and communication.

\section{Self-healing Bessel modes}
\noindent Since Bessel modes cannot be realized experimentally, a valid approximation, the Bessel-Gaussian (BG) mode, is commonly used \cite{Gori1987}. This approximation inherits from the Bessel modes the ability to self-reconstruct in amplitude, phase \cite{McGloin2003,Litvin2009}, and polarization \cite{Milione2015d,Wu2014,Li2017}, even when considering entangled photon pairs \cite{McLaren2014} or non-separable vector modes \cite{otte2018recovery,Rosales2017,Otte2018}. Mathematically, they are described by  

\begin{align}
\mathcal{J}_{\ell,k_r}(r,\varphi, z) = &\sqrt{\frac{2}{\pi}}J_{\ell}\left(\frac{z_R k_r r}{z_R-\text{i}z}\right)\exp\left(\text{i}\ell\varphi-\text{i}k_z z\right) \nonumber \\
&\cdot \exp\left(\frac{\text{i}k_r^2z w_0-2kr^2}{4(z_R-\text{i}z)}\right),
\label{eq:BGmodes}
\end{align}
where $(r,\varphi, z)$ represents the position vector in the cylindrical coordinates, $\ell$ is the azimuthal index (topological charge). Furthermore, $J_{\ell}(\cdot)$ defines a Bessel function of the first kind , $k_r$ and $k_z$ are the radial and longitudinal components of the wave number $k=\sqrt{k_{r}^2 + k_{z}^2}=2\pi/\lambda$. The last factor describes the Gaussian envelope with beam waist $w_0$ and Rayleigh range $z_R = \pi w_0^2/\lambda$ for a certain wavelength $\lambda$.

The propagation distance over which the BG modes approximate a non-diffracting mode is given by $z_{\text{max}} = 2\pi w_0/\lambda k_r$. In the presence of an obstruction of radius $R$ inserted within the non-diffracting distance, a shadow region of length $z_{\text{min}} \approx 2\pi R/k_r \lambda$ is formed\,\cite{Bouchal1998}. The distance $z_{\text{min}}$ determines the minimum distance required for the beam to recover its original form, whereby full reconstruction is achieved at 2$z_{\text{min}}$\,\cite{McGloin2003,Litvin2009}.

We exploit this property with single photons that have non-separable polarization and OAM DoFs. By carefully selecting a $k_r$ value, we show that the information of hybrid entangled single photon encoded with a Bessel radial profile can be recovered after the shadow region of an obstruction. Traditionally hybrid modes, while still new in the communication context, have not been controlled in radial profile. Indeed, the traditional generation approaches often result in very complex radial structures \cite{sephton2016revealing}.  To control and exploit all spatial and the polarization DoFs for QKD we introduce a high-dimensional self-healing information basis constructed from non-orthogonal vector and scalar OAM BG spatial modes.

\subsection{Self-healing information basis}
\begin{figure*}[t]
\centering
\def\svgwidth{1.0\linewidth}
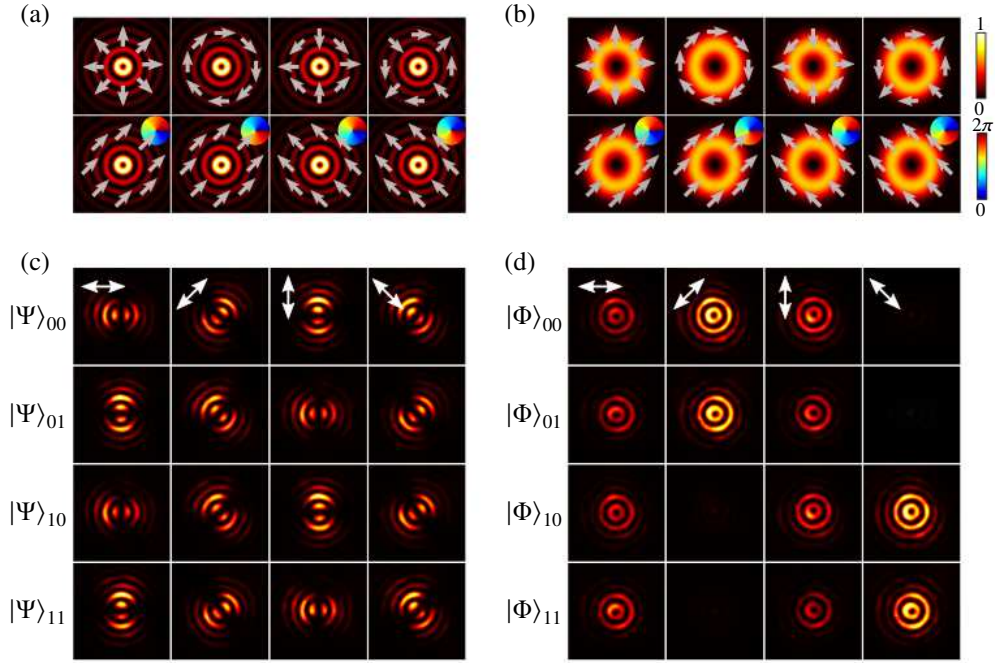
\caption{Intensity and polarization mappings of vector (first row) and scalar (second row) MUB modes with (a) BG and (b) LG radial profiles for $\ell=\pm1$. The polarization projections on the (c) vector $\ket{\Psi}$ and (d) scalar $\ket{\Phi}$ basis BG modes. The vector modes have spatially varying polarizations which consequently render the polarization and spatial DoF as non-separable. This is easily seen in the variation of the transverse spatial profile when polarization projections are performed (orientation indicated by white arrow) on the $\ket{\Psi}$ modes. In contrast, the scalar modes have separable polarization and spatial DoF hence polarization projections only cause fluctuations in the intensity of the transverse profile for the $\ket{\Phi}$ modes.} \label{fig:SpatialProfiles}
\end{figure*}
In order to demonstrate the concept we will use the well-known BB84 protocol, but stress that this may be replaced with more modern and advantageous protocols with little change to the core idea as outlined here. In the standard BB84  protocol, Alice and Bob unanimously agree on two information basis. The first basis can be arbitrarily chosen in $d$ dimensions as $\{\ket{\Psi_i}, i=1..d\}$. However, the second basis must fulfill the condition

\begin{equation}
|\braket{\Psi_i|\Phi_j}|^{2}=\frac{1}{d},
\end{equation}
making $\ket{\Psi}$ and $\ket{\Phi}$ mutually unbiased. Various QKD protocols were first implemented using polarization states, spanned by the canonical right $\ket{R}$ and left $\ket{L}$ circular polarization states constituting a two-dimensional Hilbert space, i.e.,  $\mathcal{H}_\sigma=\text{span}\{\ket{L}, \ket{R}\}$. More dimensions where later realized with the spatial DoF of photons \cite{mafu2013higher,mirhosseini2015high}, using the OAM DoF spanning the infinite dimensional space, i.e. $\mathcal{H}_\infty = \bigoplus\mathcal{H}_\ell$, such that $\mathcal{H}_{\ell}=\{\ket{\ell}, \ket{-\ell}\}$ is qubit space characterized by a topological charge $\ell\in\mathbb{Z}$.

Here, we exploit an even larger encoding state space by combining polarization and OAM, $\mathcal{H}_\infty=\bigoplus\mathcal{H}_\sigma\otimes\mathcal{H}_{\ell}$ where $\mathcal{H}_4=\mathcal{H}_\sigma\otimes\mathcal{H}_{\ell}$, is a qu-quart space spanned by the states $\{\ket{L}\ket{\ell}, \ket{R}\ket{\ell}, \ket{L}\ket{-\ell}, \ket{R}\ket{-\ell}\}$, described by the so-called higher-order Poincar\'e spheres (HOPSs) \cite{Milione2011a,holleczek2011classical}. These modes feature a coupling between the polarization and OAM DoFs, shown in Fig.~\ref{fig:SpatialProfiles}. The HOPS concept neglects the radial structure of the modes, considering only the angular momentum content, spin and orbital. Yet all modes have radial structure, shown in Fig.~\ref{fig:SpatialProfiles} (a) for BG and (b) for LG profiles. We wish to create a basis of orthogonal non-separable vector BG modes together with their MUBs for our single photon states.

Without loss of generality, we choose a mode basis on the $\mathcal{H}_4$ subspace with $\ell=\pm1$ as our example. Our encoding basis is constructed as follows: we define the radial profile $\mathcal{J}_{\ell,k_r}(r)$  representing the radial component of the BG mode in Eq. (\ref{eq:BGmodes}). Our first mode set is comprised of a self-healing vector BG mode basis, mapped as

\begin{eqnarray}
\ket{\Psi}_{00} &=& \frac{1}{\sqrt{2}} \mathcal{J}_{\ell,k_r}(r)\big(\ket{R}\ket{\ell}+\ket{L}\ket{-\ell}\big),\\
\ket{\Psi}_{01} &=& \frac{1}{\sqrt{2}}\mathcal{J}_{\ell,k_r}(r)\big(\ket{R}\ket{\ell}-\ket{L}\ket{-\ell}\big),\\
\ket{\Psi}_{10} &=& \frac{1}{\sqrt{2}}\mathcal{J}_{\ell,k_r}(r)\big(\ket{L}\ket{\ell}+\ket{R}\ket{-\ell}\big),\\
\ket{\Psi}_{11} &=&\frac{1}{\sqrt{2}}\mathcal{J}_{\ell,k_r} (r)\big(\ket{L}\ket{\ell}-\ket{R}\ket{-\ell}\big),
\end{eqnarray}
\noindent with some example polarization projections shown in Fig. \ref{fig:SpatialProfiles} (c). The set of MUB modes is given by

\begin{eqnarray}
\ket{\Phi}_{00} &=&\mathcal{J}_{\ell,k_r}(r)\ket{D}\ket{-\ell}, \\
\ket{\Phi}_{01} &=& \mathcal{J}_{\ell, k_r} (r)\ket{D}\ket{\ell}, \\
\ket{\Phi}_{10} &=&\mathcal{J}_{\ell,k_r}(r)\ket{A}\ket{-\ell}, \\
\ket{\Phi}_{11} &=& \mathcal{J}_{\ell,k_r}(r)\ket{A}\ket{\ell}, 
\end{eqnarray}

\noindent where $D$ and $A$ are the diagonal and anti-diagonal polarization states (see Fig.~\ref{fig:SpatialProfiles} (d) for polarization projections). The set $\ket{\Psi}_{ij}$ and $\ket{\Phi}_{ij}$ are mutually unbiased and, therefore, form a reputable information basis for QKD in high dimensions.

As a point of comparison to the self-healing properties of the non-diffracting modes, we make use also of a similar alphabet but projecting the heralding photon onto a Gaussian mode, obtaining a helical mode in the other photon after traversing a spin-to-orbital angular momentum converter \cite{Marrucci2006}. We will refer to this as a LG mode in the remainder of the manuscript.

\section{Methods}
\subsection{Single photon heralding}

\begin{figure*}[ht!]
\centering
\def\svgwidth{1\linewidth}
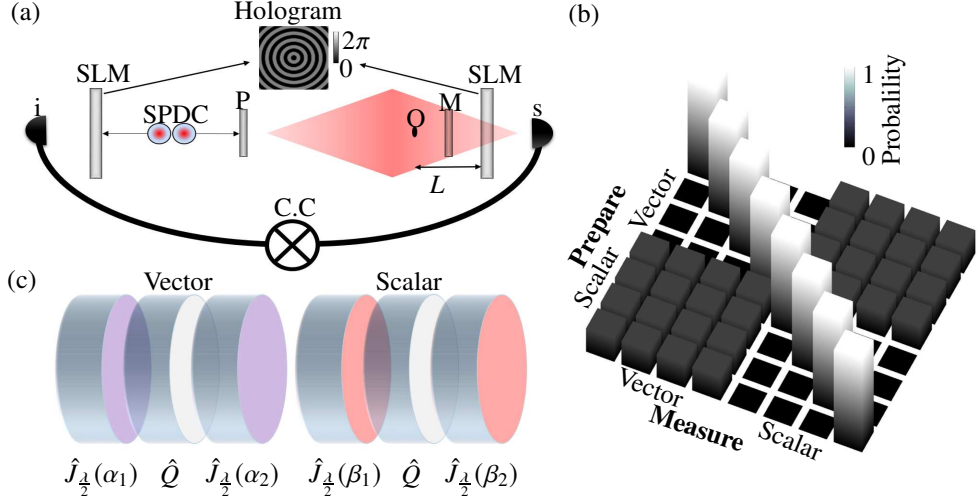
\caption{ (a) Conceptual drawing of the QKD with self-healing BG modes. The SLMs post-select the self-healing BG radial profile from the SPDC source. The prepare (P) and measure (M) optics modulate and demodulate the OAM and polarization DoF of the heralded photon. The physical obstruction (O) is placed at a distance $L$ from the rightmost SLM, which decodes the radial information of Bob's photon. The optics are within $z_\text{max}=54$ cm distance of the BG modes depicted as the rhombus shape. The propagation of the post-selected BG mode can be determined via back-projection. (b) Numerical scattering probability matrix for the vector and scalar modes sets in free-space. The channels correspond to the probabilities $|C_{ij}|^2$ calculated from Eq. (\ref{OverlapVMSM}). (c) Optical elements required by Alice and Bob to prepare and measure the spin-coupled states of the heralded photons (cf. Table \ref{table:table2}).}
\label{fig:Concept}
\end{figure*}

Heralded photon sources have been used as a means of producing single photons in QKD \cite{eisaman2011invited}. In this process, the heralded photon conditions the existence of its correlated twin. Moreover, the statistics of the heralded photon have low multi-photon probabilities {which can be further remedied by using decoy states \cite{wang2008exp}. 

Here, we herald a single photon via SPDC where a high frequency photon ($\lambda$ = 405 nm) is absorbed with low probability in a nonlinear crystal, generating a signal ($s$) and idler ($i$) correlated paired photons at $\lambda$ = 810 nm. In the case of a collinear emission of $s$ and $i$, the probability amplitude of detecting mode functions $\ket{m}_s$ and $\ket{m}_i$, respectively, is given by~\cite{zhang2014simulating}

\begin{equation}
c_{s,i}=\int\int m^* _s (\textbf{x}) m^* _i (\textbf{x}) m_p(\textbf{x}) d^2 x \label{OverLap},   
\end{equation}
where $m_p(\textbf{x})$ is the field profile of the pump ($p$) beam which best approximates the phase-matching condition in the thin crystal limit; the Rayleigh range of the pump beam is much larger than the crystal length. The probabilities amplitudes $c_{s,i}$ can be calculated using the Bessel basis,
\begin{equation}
m_{s,i}(r, \varphi) = \mathcal{J}_{\ell_{s,i},k_r}(r) \hspace{0.1 cm} \text{exp}(\text{i}\ell_{s,i}\varphi), 
\end{equation}
where $\text{exp}(\text{i}\ell\varphi)$ corresponds to the characteristic azimuthal phase mapping onto the state vector $\ket{\ell}$.
Taking into account a SPDC type-I process and a Gaussian pump beam, the quantum state used to encode and decode the shared key can be written in the Bessel basis as
\begin{equation}
\ket{\Psi}_{AB} = \sum c_{\ell, k_{r,1}, k_{r,2}} \ket{\ell, k_{r,1}}_s \ket{-\ell, k_{r,2}}_i \ket{H}_s\ket{H}_i,
\end{equation}
being $\ket{\ell, k_{r}}_s\sim J_{\ell,k_r}( r)\ket{\ell}$ and $H$ the horizontal polarization state. The probability amplitudes $c_{\ell, k_{r,1}, k_{r,2}}$ can be calculated using the overlap integral in Eq. (\ref{OverLap}). Experimentally $|c_{\ell, k_{r,1}, k_{r,2}}|^2$ is proportional to the probability of detecting a coincidence when the state $\ket{\ell, k_{r,1}}_s\ket{-\ell, k_{r,2}}_i$ is selected. Coincidences are optimal when $|k_{r,1}|$ and $|k_{r,2}|$ are equivalent.

In this experiment, the idler photon ($i$) is projected into the state $\ket{0, k_{r}}_i$, heralding only the signal photons ($s$) with the same spatial state $\ket{0, k_{r}}_s$, as can be seen in the sketch of Fig.~\ref{fig:Concept}(a). Therefore, a prepare-measure protocol can be carried out by using the same $s$ photon. In other-words, Alice remotely prepares her single photon with a desired radial profile from the SPDC before encoding the polarization and OAM information. 

\subsection{Spatial profile post-selection}
\noindent Spatial light modulators (SLMs) are a ubiquitous tool for generating and detecting spatial modes \cite{SPIEbook,Forbes2016}. We exploit their on-demand dynamic modulation via computer generated holograms to post-select  the spatial profiles of our desired modes (see hologram inset in Fig.~\ref{fig:Concept}(a)). For the detection of BG modes, we choose a binary Bessel function as phase-only hologram, defined by the transmission function
\begin{equation}
T(r,\varphi) = \text{sign}\lbrace J_{\ell}(k_r r)\rbrace \exp (\text{i}\ell\varphi), \label{eq:BinaryBessel}
\end{equation}
with the sign function $\text{sign}\lbrace\cdot \rbrace$ \cite{Turunen1988,Cottrell2007}. Classically, this approach has the advantage of generating a BG beam immediately after the SLM and, reciprocally, detects the mode efficiently \cite{McLaren2014}. Importantly, a blazed grating is used to encode the hologram, with the desired mode being detected in the first diffraction order\cite{Davis1999} and spatial filtered with a single mode fiber (SMF).

Here, we set $k_r = 18\,\mbox{rad/mm}$ and $\ell =0$ for the fundamental Bessel mode  and, conversely, $k_r=0$ to eliminate the multi-ringed Bessel structure.

\subsection{Mode generation and detection}
Liquid crystals $q$-plates represent a convenient and versatile way to engineer several types of vector beams \cite{Cardano2012}. In our setup, vector and scalar modes, described in Fig. \ref{fig:SpatialProfiles}, are either generated or detected, at Alice and Bob's prepare (P) and measure (M) stations in Fig. \ref{fig:Concept} (a), by letting signal photons pass through a combination of these devices and standard wave plates (see Fig.~\ref{fig:Concept} (c)).  A $q$-plate consists of a thin layer of liquid crystals (sandwiched between glass plates) whose optic axes are arranged so that they form a singular pattern with topological charge $q$ \cite{Marrucci2006}. By adjusting the voltage applied to the plate it is possible to tune its retardation to the optimal value $\delta=\pi$ \cite{piccirillo2010photon}. In such a configuration indeed the plate behaves like a standard half-wave plate (with an inhomogeneous orientation of its fast axis) and can be used to change the OAM of circularly polarized light by $\pm 2q$, depending on the associated handedness being left or right, respectively. In the Jones matrix formalism, the $q$-plate is represented by the operator
 
 \begin{equation}
  \hat{Q} =  \begin{pmatrix} 
  			\mathrm{cos}(2q\varphi) & 	\mathrm{sin}(2q\varphi) \\ 
 			\mathrm{sin}(2q\varphi) & 	-\mathrm{cos}(2q\varphi),
 		   \end{pmatrix}
           \label{eq:q-plateMatrix}
 \end{equation}
 
\noindent where $\varphi$ is the azimuthal coordinate. The matrix is then written in the following linear basis $\lbrace\ket{H}=\begin{pmatrix}1 \\ 0 \end{pmatrix}, \ket{V}=\begin{pmatrix}0 \\ 1 \end{pmatrix}\rbrace$. In our experiment we use $q$-plates with $q=1/2$, and half-wave ($\frac{\lambda}{2}$) as well as quarter-wave ($\frac{\lambda}{4}$) plates for polarization control, represented by the Jones matrices
 
 \begin{align}
 \hat{J}_{\frac{\lambda}{2} } (\theta)=  \begin{pmatrix} 
 \mathrm{cos}(2\theta) & 	\mathrm{sin}(2\theta) \\ 
 \mathrm{sin}(2\theta) & 	-\mathrm{cos}(2\theta) \end{pmatrix} ,
 \end{align}
 
 \noindent and 
 
  \begin{align}
\hat{J}_{\frac{\lambda}{4} } (\theta)=  \begin{pmatrix} 
 \mathrm{cos}^2(\theta) + \text{i}\mathrm{sin}^2(\theta)  & 	(1-\text{i})\,\mathrm{sin}(\theta)\mathrm{cos}(\theta) \\ 
 (1-\text{i})\,\mathrm{sin}(\theta) \mathrm{cos}(\theta) & \mathrm{sin}^2(\theta) + \text{i}\mathrm{cos}^2(\theta) 
 \end{pmatrix}.
 \end{align}
\noindent Here, $\theta$ represents the rotation angle of the wave plates fast axis with respect to the horizontal polarization. The operator associated with the generation of the vector mode is

\begin{equation}
\hat{V}(\alpha_1,\alpha_2)= \hat{J}_{\frac{\lambda}{2}} (\alpha_2) \hat{Q} \hat{J}_{\frac{\lambda}{2}}(\alpha_1)\hat{P}_{H},
\end{equation} 

\noindent  where $\alpha_1$ and $\alpha_2$ are the rotation angles for the half-wave plates and $\hat{P}_{H}= \begin{pmatrix} 1  & 	0 \\ 0 &0 \end{pmatrix}$ represents the operator for a horizontal linear polarizer. Similarly, the operator for the scalar modes is
\begin{equation}
\hat{S}(\beta_1,\beta_2) = \hat{J}_{\frac{\lambda}{4}} (\beta_2) \hat{Q} \hat{J}_{\frac{\lambda}{4}}(\beta_1)\hat{P}_{H},
\end{equation}

\noindent where $\beta_1$ and $\beta_2$ are the rotation angles for the quarter-wave plates.
 
Let the set $\mathcal{M}_1=\{\hat{V}_i \,|\, \hat{V}_i\rightarrow\ket{\Psi_i}, i=1..4 \} $ be associated with the generation of vector modes from $\hat{V}(\alpha_1,\alpha_2)$, and $\mathcal{M}_2=\{\hat{S}_j \, | \, \hat{S}_j\rightarrow\ket{\Phi_j} \,, \, j=1..4 \} $ for the scalar modes from $\hat{S}(\beta_1,\beta_2)$. The orientation of the angles required to obtain them is given in Table \ref{table:table2} for the vector and scalar modes (see also schematics of wave plates arrangement in Fig. \ref{fig:Concept} (c)).

\begin{table}[h]
\caption{Generation of vector and scalar modes from a horizontally polarized BG mode ($\ell=0$) at the input. The angles $\alpha_{1,2}$ and $\beta_{1,2}$ are defined with respect to the horizontal polarization. For each $\hat{V}_i$ and $\hat{S}_i$ we present the angles needed to perform the mapping of $\mathcal{M}_1 \rightarrow \{\ket{\Psi_i}\}$ and $\mathcal{M}_2 \rightarrow \{\ket{\Phi_i}\}$ with a one-to-one correspondence.}
	\centering
	
\begin{tabular}{|c|c|c|c|c|c|}
		\hline
		\multicolumn{3}{|c|}{Vector, $\hat{V}(\alpha_1$, $\alpha_2$)}& \multicolumn{3}{|c|}{Scalar, $\hat{S}(\beta_{1}, \beta{2})$}\\
		\hline \hline 
		Operator &$ \hat{J}_{\frac{\lambda}{2} } (\alpha_1) $ & $ \hat{J}_{\frac{\lambda}{2} } (\alpha_2)$ & Operator& $ \hat{J}_{\frac{\lambda}{4}} (\beta_1)$  & $ \hat{J}_{\frac{\lambda}{4} } (\beta_2)$ \\
		\hline 
		$\hat{V}_1$ &  0 & --&$\hat{S}_1$ & $-\pi/4$  & $0$   \\ 
		\hline 
		$\hat{V}_2$& $\pi/4$  & --&$\hat{S}_{2}$ & $\pi/4$& $\pi/2$  \\ 
		\hline 
		$\hat{V}_3$& 0  & 0 &$\hat{S}_{3}$ & $-\pi/4$ & $\pi/2$\\ 
		\hline 
		$\hat{V}_4$&  $\pi/4$  & $0$ &	$\hat{S}_{4}$ & $\pi/4 $ & $0$ \\ 
		\hline
	\end{tabular}
	\label{table:table2}
\end{table}

\subsection{Scattering probability }

Let  $\hat{A}_i, \hat{B}_j \in \mathcal{M}_1\cup\mathcal{M}_2$ represent operators selected by Alice and Bob, respectively. Alice first obtains a heralded photon from the SPDC with the input state $\ket{\psi_{\text{in}}}= \mathcal{J}_{0,k_r}\ket{H}$. Then, Alice prepares the photon in a desired state from the MUB with

\begin{equation}
\ket{a_{i}} =\hat{A}_i \mathcal{J}_{0,k_r} (r) \ket{H},
\end{equation}

\noindent and Bob similarly measures the state
\begin{equation}
\ket{b_{j}}  =\hat{B}_j  \mathcal{J}_{0,k_r} (r) \ket{H}.
\end{equation}

\noindent The probability amplitude of Bob's detection is  
\begin{align}
C_{ij}&= \braket{b_j|a_i} = \int_0^{2\pi} \int_0^\infty  \bra{H}\mathcal{J}^*_{0,k_r} (r) \hat{B}^\dagger_j\hat{A}_i \mathcal{J}_{0,k_r} (r) \ket{H} r dr d\phi,\label{OverlapVMSM}
\end{align} 

\noindent while the corresponding detection probabilities, $|C_{ij}|^2$, are presented in Fig. \ref{fig:Concept} (b).

\section{Experimental set-up}
Figure \ref{fig:Set-Up} is a schematic representation of our experimental setup. The continuous-wave pump laser (Cobalt MLD diode laser, $\lambda = 405\,\mbox{nm}$) was spatially filtered to deliver 40 mW of average power in a Gaussian beam of $w_0 \approx 170$ $\mu$m at the crystal (2-mm-long PPKTP nonlinear crystal), generating two lower-frequency photons by means of a type-I spontaneous parametric down-conversion (SPDC) process. By virtue of this, the signal and idler photons had the same wavelength ($\lambda = 810\,\mbox{nm}$) and polarization (horizontal).

\begin{figure*}[t!]
\centering
\def\svgwidth{0.9\linewidth}
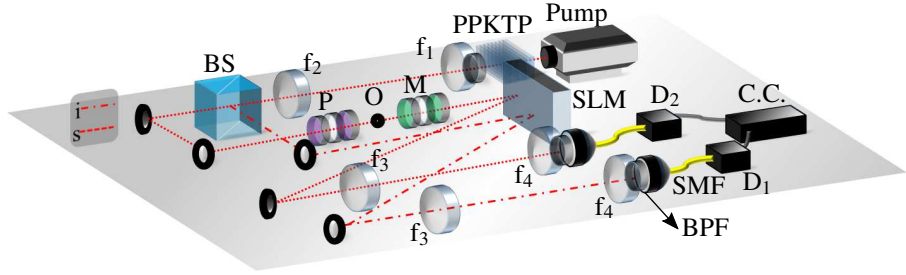
\caption{Experimental setup for the self-healing QKD. Pump: $\lambda = 405\,\mbox{nm}$ (Cobalt, MLD laser diode); f: Fourier lenses of focal length f$_{1,2,3\&4}$ = 100 mm, 750 mm, 500 mm, 2 mm, respectively; PPKTP: periodically poled potassium titanyl phosphate (nonlinear crystal); BS: 50:50 beam splitter; s and i: signal and idler photon paths; P: preparation of the state (Alice); O: variable sized obstacle; M: measurement of the state (Bob); SLM: spatial light modulator (Pluto, Holoeye); BPF: band-pass filter; SMF: single mode fiber; D$_{1\&2}$: single photon detectors (Perkin Elmer); C.C.: coincidence electronics.\label{fig:Set-Up}}
\end{figure*}

The two correlated photons, signal and idler, were spatially separated by a 50:50 beam splitter (BS), with the idler photon projected into a Bessel state of 0 OAM, thus heralding a zero-order Bessel photon in the signal arm for the prepare-measure BB84 protocol. The signal photon traversed the preparation stage (P) where Alice could prepare a vector or scalar state from the MUB alphabet using elements detailed in Fig.~\ref{fig:Concept} (c). The signal photon was then propagated in free-space with an obstacle of variable size placed within the non-diffracting distance. This mimics a line-of-sight quantum channel.  In our experiment we used the spatial light modulators (SLMs) to post-select a wave number of $k_r = 18\,\mbox{rad/mm}$, thus realising a non-diffracting distance of $z_{\text{max}} = 54$ cm.  These values where verified by classical back-projection through the system \cite{mclaren2013two}.  The state measurement (M) was implemented after the obstacle by Bob. The SLM acted both as a horizontal polarization filter and as a post-selecting filter for the radial wave number. To conclude the heralding experiment, both photons were spectrally filtered by band-pass filters (10 nm bandwidth at full-width at half-maximum) and coupled with single mode fibers to single photon detectors (D$_{1\&2}$; Perkin-Elmer), with the output pulses  synchronized with a coincidence counter (C.C.), discarding also the cases where the two photons exit the same output port from the BS.

\subsection{Procedure and analysis}
We measured the scattering matrix for the BG and, for comparison reasons, the LG profiles under three conditions: (FS) in free-space; (R$_1$) with a $600$ $\mu$m radius obstruction placed strategically such that the complete decoding is performed after $L>z_{\text{min}}$ ($L$: distance between obstruction and decoding SLM); and (R$_2$) with a $800$ $\mu$m radius obstruction, placed at the same position.  In the (R$_2$) the shadow region overlaps the detection system ($L<z_{\text{min}}$) so that the mode is not able to self-reconstruct completely before being detected. We measure the quantum bit error rate (QBER) in each of these cases and computed the mutual information between Alice and Bob in $d=4$ dimensions by \cite{cerf2002security}
\begin{equation}
I_{AB}=\log_{2}(d) + (1-e)\log_{2}(1-e)+ (e)\log_{2}\left(\frac{e}{d-1}\right).\label{eq:MutualInfo}
\end{equation}
Here, $e$ denotes the QBER. Lastly, we measured the practical secure key rate per signal state emitted by Alice, using the Gottesman-Lo-L\"{u}tkenhaus-Preskill (GLLP) method \cite{gottesman2004security,schiavon2016heralded} for practical implementations with BB84 states, given by
\begin{equation}
R_{\Delta} = Q_{\mu} \left( (1-\Delta) \left( 1 - H_d\left(\frac{e}{1-\Delta}\right)\right) - f_{\text{EC}}H_d(e) \right) \label{eq:KeyRate},
\end{equation}
where $H_d(\cdot)$ is the high-dimensional Shannon entropy and $f_{\text{EC}}$ is a factor that accounts for error correction and is nominally $f_{\text{EC}}= 1.2$ for error correction systems that are currently in practice.

 \begin{figure*}[t!]
\centering
\def\svgwidth{0.9\linewidth}
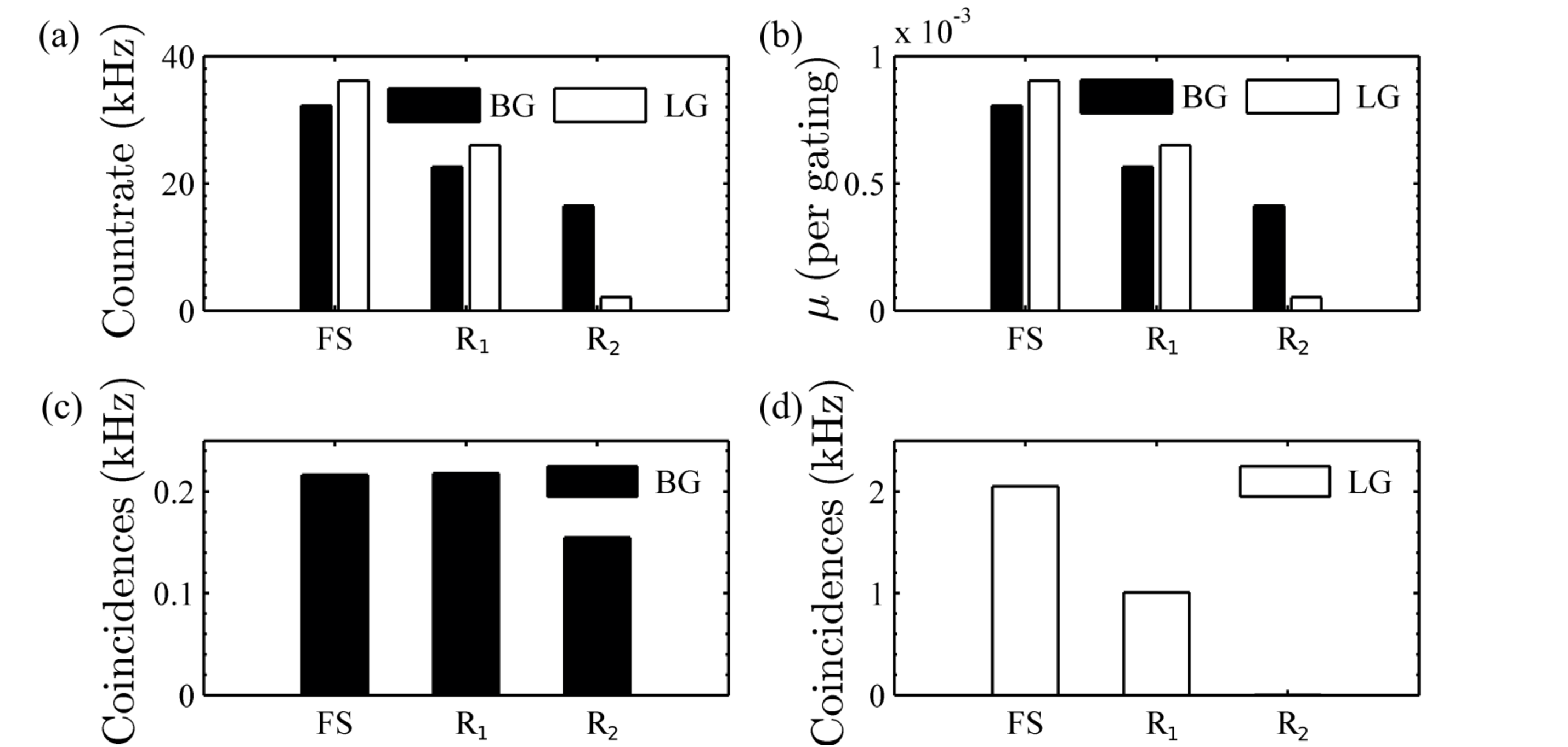
 \caption{(a) Measured photon count rates and (b) average photon number ($\mu$) per-gating window of 25 ns in free-space (FS) and the two obstructions (R$_1$ = 600 $\mu$m and R$_2$ = 800 $\mu$m) for the radially polarized mode $\ket{\psi}_{00}$. (c) and (d) show coincidence rates with the same obstructions for the BG and LG radial profiles, respectively. The BG count rate is lower for smaller obstructions due to the high $k_r$ hologram on the SLM \cite{mclaren2013two}.
 \label{fig:CountsAnalysis}}
\end{figure*}

The photon gain is defined as $Q_\mu = \sum_n Y_n P_n(\mu)$ (in the orders of $10^{-4}$ for our experiment), where $Y_n$ is the $n$-th photon yield while $P_n$ is the probability distribution over $n$ with respect to the average photon number $\mu$, following sub-Poisson statistics for heralded photons produced from a SPDC source \cite{schiavon2016heralded}. $Y_n$ can be calculated from the background rate, $p_D = 2.5 \times 10^{-6}$ photons per gating window (25 ns), and $n$-signal detection efficiency $\eta_n$:
\begin{align}
Y_n = \eta_n+ p_D(1 -\eta_n),
\end{align}
\noindent where the $n$-signal detection efficiency $\eta_n$ is given by
\begin{align}
\eta_n=1-(1-\eta)^n.
\end{align}

\noindent Here $\eta=\eta_{d}t_B$ is the transmission probability of each photon state with $\eta=0.45 \times 0.8$ for Bob's detection (when accounting for the SLM grating). Furthermore, $\Delta$ is the multi-photon rate computed as $(1-P_0-P_1)/Q_{\mu}$ \cite{schiavon2016heralded} where $P_{0,1}$ are the vacuum and single photon emission probabilities, respectively. The term ($1-\Delta$) accounts for photon splitting attacks \cite{schiavon2016heralded}. In our experiment, we measured the photon intensities for every obstruction from the photon detection rates of the obstructed photon and deduced $P_1$ and $P_0$ assuming a thermal statistics of the heralded photon. We point out that it may be necessary to implement decoy states with a heralded source to ensure security against multi photon states owing to the thermal nature of the reduced photon state of SPDC correlated pairs  \cite{schiavon2016heralded, eisaman2011invited}.

\section{Results and Discussion}

We performed the aforementioned experiment in four dimensions using heralded single photons with either a heralded LG mode or BG mode for the radial spatial profile, and compare their performance under the influence of varying sized obstructions.

\subsection{Experimental results}
The photon count-rates, mean-photon counts (per gating window) and coincidence-rates are presented in Fig.~\ref{fig:CountsAnalysis} (a) and (b), for the $\ket{\Psi}_{00}$ input state.  As shown, the photon count rates decay for both the BG and LG radial profiles, however, more so for LG profile under the R$_2$ obstruction. The coincidences rates are recovered for the BG mode (Fig.~\ref{fig:CountsAnalysis} (c)) under the R$_1$ obstruction since L$>z_{\mbox{min}}$ (detection is performed outside the shadow region of the obstruction). Further, the BG mode still demonstrates less decay for R$_2$ obstructed even when the mode has not reconstructed (since $L<z_{\mbox{min}}$), as compared to LG (Fig. \ref{fig:CountsAnalysis} (d)), where the coincidence rate is seen to completely decay. 

\begin{figure*}[t!]
\centering
\def\svgwidth{0.9\linewidth}
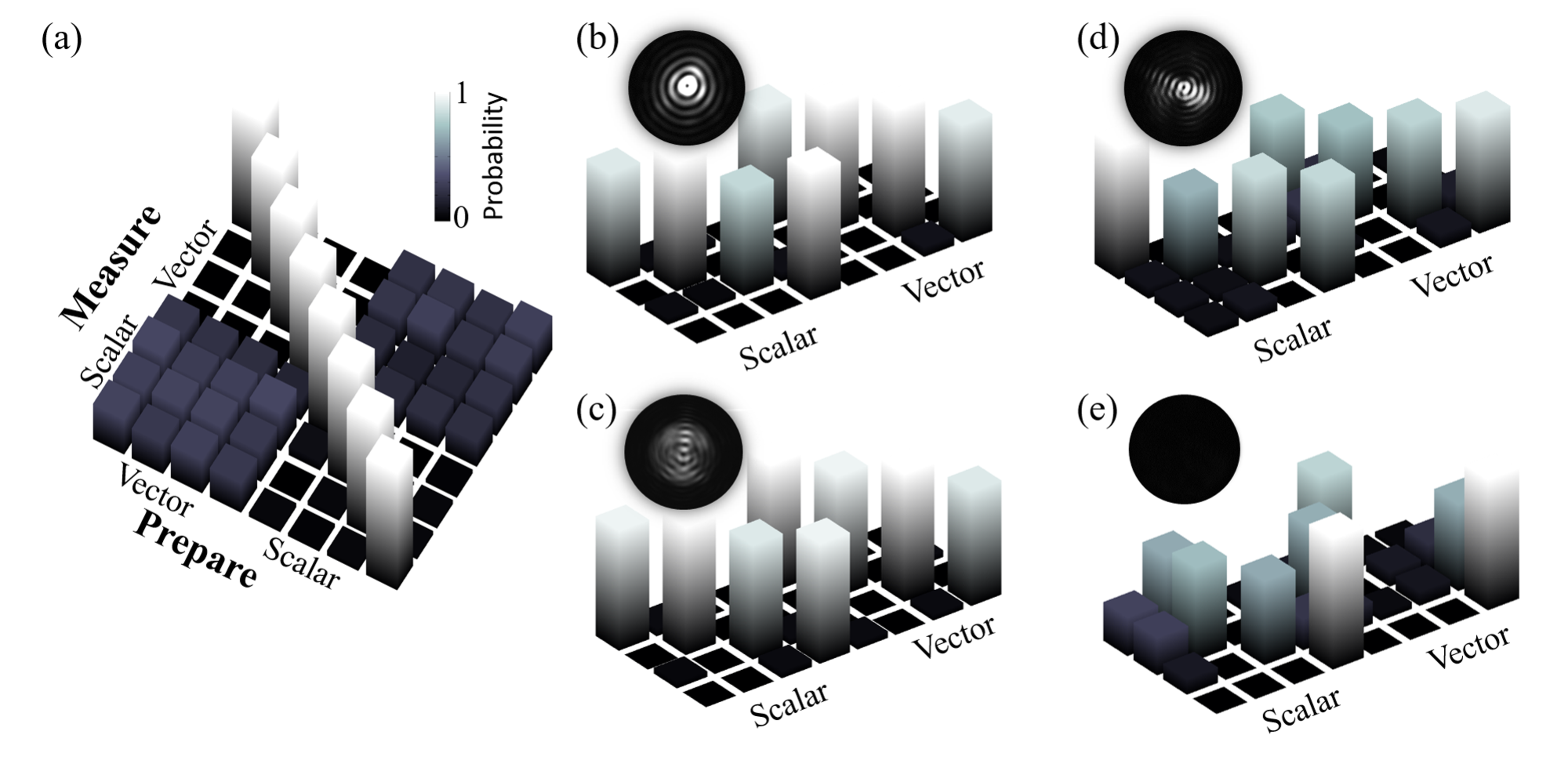
\caption{Crosstalk (scattering) matrix for  vector and scalar modes in (a) (I) free-space having post-selected in a BG radial profile. The vector and scalar measured probabilities with the first obstruction (II) having a radius R$_1 = 600$ $\mu$m ($L>z_{\text{min}}$) when taking into account (b) BG  and (c) LG radial profiles. Measured probabilities with (III) an obstruction of R$_2= 800$ $\mu$m ($L<z_{\text{min}}$) when taking into account (d) BG and (e) LG radially profiled single photons. \label{fig:Results}}
\end{figure*}

Next, we present the measured detection probability matrices for three tested cases in Fig.~\ref{fig:Results} using our high-dimensional information basis. In the free-space case, we measure QBERs of $e=0.04\pm0.004$ for the  BG and LG spatial profiles (see Fig.~\ref{fig:Results} (a) and Table \ref{table:table4}). We compute a mutual information of I$_{AB}=1.69$ bits/photon and a secure key rate of $R_\Delta/Q_\mu = 1.32$ bits/s per photon for both radial profiles. 

\begin{figure}[t!]
\centering
\def\svgwidth{0.9\linewidth}
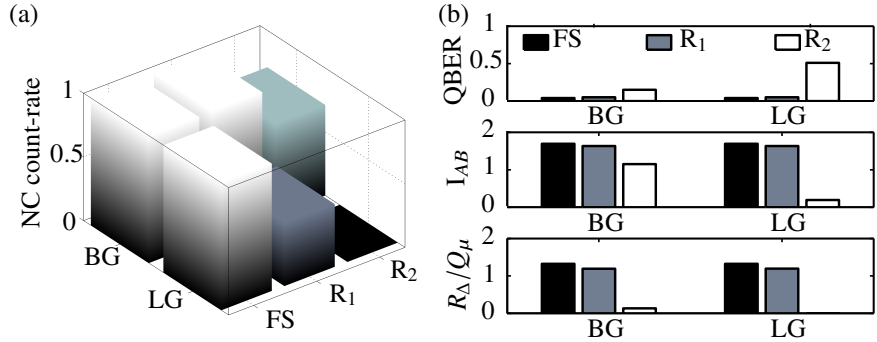
\caption{(a) Experimental normalized coincidence (NC) count-rate for the BG and LG MUB  for free-space (FS) and the two obstructions (R$_1$ = 600 $\mu$m and R$_2$ = 800 $\mu$m) on the radially polarized mode $\ket{\psi}_{00}$. (b) The QBER, mutual information ($I_{AB}$) and key rate ($R_{\Delta}/Q_{\mu}$) for the BG and LG modes with no perturbation and under the two tested obstructions are shown.} \label{fig:Security analysis}
\end{figure}

Under the perturbation of the R$_1 = 600$  $\mu$m obstruction ($0.53\times$ the beam waist of the down converted photon), we measure a QBER of $e=0.05$ for both spatial profiles, indicative of information retention, i.e. high fidelity. The intensity fields from the back-projected classical beam (see insets of Fig.~\ref{fig:Results} (b) and (c)), show self-healing of the BG mode at the SLM plane (see Fig.~\ref{fig:Results} (b)), although the LG is not completely reconstructed (see Fig.~\ref{fig:Results} (c)). The photons encoded with the LG profile may have a large component of the input mode which is undisturbed in polarization and phase. Furthermore, the coincidence counts decreases to $49\%$ for the LG profile relative to the counts in free-space, as highlighted in Fig.~\ref{fig:Security analysis} (a). In comparison, the BG modes show resilience thanks to the multiple concentric rings \cite{mendoza2015laguerre}.

Lastly, we investigate the security when the R$_2= 800$ $\mu$m ($0.71\times$ the beam waist of the down converted photon) obstruction is used. Remarkably, as illustrated in Fig.~\ref{fig:Security analysis} (a), the signal decreased by almost four orders of magnitude, remaining only the $0.07\%$ of the signal for the LG set, but up to $71\%$ for the BG self-healing mode set, owing to an earlier reconstruction of the BG radial profile in comparison to the LG radial profile. Based on the measurement results shown in Fig.~\ref{fig:Results} (d) and (e), we determine a QBER of $e=0.15\pm0.01$ and $e=0.51\pm0.00$ for the BG and LG modes, respectively. The mutual information (I$_{AB}$) and secure key rates are higher for the BG basis than the LG, even though the BG MUB has not fully reconstructed (see Fig.~\ref{fig:Security analysis} (b)). Table \ref{table:table4} shows a summary of the measured security parameters for the BG and LG mode sets.

\begin{table}[h!]
	\caption{ Measured security parameters for the self-healing BG (LG) modes. NC represents the normalized coincidence counts. The normalization was performed with respect to the counts obtained from the free-space measurements.}

\hspace{-0.1 cm}\begin{tabular}{|c|c|c|c|}
	
\hline
	\multicolumn{4}{|c|}{BG (LG) modes}  \\
	\hline 
        \hline
	&Free-space & R$_1=600\,\mbox{\textmu m}$ & R$_2=800\,\mbox{\textmu m}$\\
     \hline
	QBER&$0.04\pm0.01$ ($0.04\pm0.01$) & $0.05\pm0.02$ ($0.05\pm0.03$) & $0.15\pm0.01$  ($0.51\pm0.00$)\\ 
	\hline 
	$I_{AB}$ &$1.69\pm0.06$ ($1.69\pm0.03$) & $1.63\pm0.1$ ($1.63\pm0.02$) &$1.15\pm04$ ($0.19\pm.004$)\\ 
	\hline
	$\Delta$ & $1.60$ $10^{-3}$ ($1.80$ $10^{-3}$) & $1.10$ $10^{-3}$ ($1.30$ $10^{-3}$) &$0.73$ $10^-3$ ($0.04$ $10^-3$)\\  		
    \hline
	$\frac{R_\Delta}{Q_{\mu}}$ & $1.32\pm0.06$ ($1.32\pm0.03$) & $1.19\pm0.1$ ($1.19\pm0.02$) & $0.13\pm04$ ($0.01\pm0.00$) \\ 
\hline 
\end{tabular}
\label{table:table4}
\end{table}

\subsection{Discussion}

We have presented a \textit{proof-of-concept} experiment highlighting the importance of structuring photons in the complete spatial mode state.  Here we have demonstrated the advantage when doing so with BG spatial modes for obstacle-tolerant QKD.  Further, we have employed hybrid spin-orbital states to access high dimensions, with the spin-orbit states used to encode the information and the radial mode used to ameliorate perturbations in the form of obstructions. Our scheme shows that with high-dimensional encoding and self-reconstruction, high information transmission rates are still achievable even in the presence of absorbing obstructions that perturb the traverse extent of the quantum channel. Our scheme exploits the radial DoF which has previously not been explored in HD QKD implementations with spatial modes. In our experiment the propagation length was tailored for laboratory implementation, but could be extended for practical long distance links as has been done at the classical level with scalar Bessel beams \cite{ahmed2016mode}.  Doing so would likely increase the beam size as well as reduce the cone angle.  In a realistic channel the obstruction could range in scale from the very small, e.g., dust particles in dry environments, to the very large, perhaps birds, and may even be in the transmitter or receiver itself, e.g., conventional mirror telescope designs that block part of the incoming light. To mimic this range in scale we have used obstructions that range in relative size to the mode from 0 (free-space) to $0.7 \times$ (very large). We have also used a very difficult high $k_r$ value of 18 rad/mm, returning meter scale distances (54 cm in our case) for a beam radius in the order of 100s of $\mu$m.  Thus links in the kilometer range could be produced with modest cm scale beams, or the heralding efficiency could be dramatically increased by lowering $k_r$ \cite{mclaren2013two} and instead increasing the beam size.  These design trade-offs are afforded to the user by the use of BG modes over LG modes. 

In free-space a common problem is phase distortions, such as turbulence. Here BG modes do not show complete reconstruction \cite{mphuthi2018bessel}, nor does the hybrid combination add value \cite{Cox2016}, but classical studies have suggested that perhaps such modes may be resilient to beam wander due to turbulence \cite{yuan2017beam}.  This is yet to be tested in the quantum regime.  We predict that the ability to tailor both the size and $k_r$ to achieve a desired distance may assist in keeping the beam size below the Fried scale.

In cases where the BG adds no advantage the radial mode should still be tailored correctly to a more appropriate choice.  In this sense this study highlights the general case for complete control of the DoFs of the state for QKD, using BG modes as an example.

We also stress that although there are reported benefits with  HD encoding, not all commonly used protocols have been generalized to high dimensions, for example, the SARG04 protocol \cite{scarani2004quantum} which is designed for robustness against the photon number splitting attacks or the B92 protocol which is a simpler version of the BB84 protocol \cite{singh2014quantum}, hence newer protocols such as the Round-Robin Differential-Phase-Shift are the subject of ongoing development in the context of spatial modes \cite{bouchard2018round}. Importantly, there may be further improvements of our work by implementing our selection of modes with decoy states which has proven invaluable for HD QKD in both free-space and fiber \cite{etcheverry2013quantum, canas2017high} and could be of higher value if implemented with heralded sources \cite{schiavon2016heralded}. Although the scheme we present is filter based, i.e. filtering states one at a time, the experiment can be performed robustly and more efficiently using a deterministic detector for spin-orbit coupled states, sorting the modes in position \cite{Ndagano2018}. This ensures high detection rates. Obtaining high switching between modes during generation would require fast modulators which is a serious experimental challenge when implementing HD QKD \cite{diamanti2016practical}.

\section{Conclusion}

The self-healing property of the Bessel-Gaussian modes opens an important research field, being able to securely share the cryptographic key despite any possible obstruction partially blocking the quantum channel. We have shown in this manuscript the experimental results of the scattering probabilities, mutual information and secret key rates in a prepare-measure protocol, comparing two different modes forming the QKD quantum state alphabet: Bessel-Gaussian (BG) and Laguerre-Gaussian (LG). Our results clearly show lower quantum bit error rate (QBER) by using BG modes when transmitting the shared key through a mostly blocked quantum channel. Concretely, we measured a QBER of $0.15\pm0.01$ and $0.51\pm0.00$ for the BG and LG modes, respectively. Furthermore, when almost completely blocking the channel, the mutual information for the BG modes only drops due to the increase of the noise with respect of the signal. The quantum state information can be reconstructed even when having barely any photons after the obstacle.

\section*{Acknowledgments}
The authors express their gratitude to Lorenzo Marrucci and Bruno Piccirillo for providing the $q$-plates. I.N., E.O., A.V. and F.C. acknowledge financial support from the Department of Science and Technology (South Africa), the German Research Foundation (DFG; DE-486-22, TRR61), the Claude Leon Foundation and the European Research Council (ERC), under Grant No. 694683 (PHOSPhOR), respectively.



\end{document}